\documentclass[mathleft
]{an}
\usepackage{graphicx}
\usepackage{times}
\usepackage{multirow}
\begin{document}

\Yearpublication{0000}%
\Yearsubmission{2010}%
\Month{00}%
\Volume{000}%
\Issue{000}%

\title{Near-surface effects and solar-age determination}

\author{G. Do\u{g}an\inst{1}\fnmsep\thanks{\email{gulnur@phys.au.dk}\newline}
A. Bonanno \inst{2} \and J. Christensen-Dalsgaard \inst{1}}

\titlerunning{Near-surface effects and the solar age}
\authorrunning{G. Do\u{g}an et al.}
\institute{Department of Physics and Astronomy, Aarhus University,
Ny Munkegade, DK-8000, Aarhus C, Denmark \and Catania Astrophysical
Observatory, Via S.Sofia 78, 95123, Catania, Italy}

\received{April 2010} \accepted{---} \publonline{later}


\abstract{%
The dominant part of the difference between the observed and model
frequencies of the Sun can be approximated by a power law. We show
that when this empirical law is employed to correct the model
frequencies and then the small frequency separations,
$\delta\nu_{02}(n)$, are used for solar age determination, the
results are consistent with the meteoritic age (4.563 Gyr $<$ t $<$
4.576 Gyr). We present the results and compare with those obtained
by using the ratios, $r_{02}(n)$, of small to large frequency
separations.}

\maketitle

\section{Introduction}

\sloppy

It is known that there is a systematic offset between the observed
and model frequencies of the Sun. This offset increases with
increasing frequency and is shown (Kjeldsen et al.~2008) to be
fitted well with a power law as
\begin{equation}
\nu_{\rm{obs}}(n)-\nu_{\rm{best}}(n)=a\left[\frac{\nu_{\rm{obs}}(n)}{\nu_{0}}\right]^b,
\end{equation} where the power $b$ is determined to be 4.90 for the Sun. Here
$\nu_{\rm{obs}}(n)$, and $\nu_{\rm{best}}(n)$, represent the
observed, and the best model, frequencies with spherical degree
$l$=0, and radial order $n$; $\nu_{0}$ is a constant frequency
(chosen to be 3100 $\mu$Hz for the Sun), which corresponds to the
frequency for peak power in the spectrum. This difference between
the observed and calculated frequencies exists due to improper
modelling of the outer turbulent convective layers of the Sun. The
outer layers affect the high frequencies most, as the upper turning
point of the high-frequency waves are closer to the surface. Since
all stellar models are calibrated with respect to the Sun, and
since, thanks to the recent developments, we are at a stage to have
observations of individual frequencies of stars other than the Sun,
it is important to understand the effects of near-surface stellar
layers on the oscillation frequencies. These effects also influence
the small frequency separations, $\delta\nu_{l \hspace{1pt}
l+2}(n)=\nu_{nl}-\nu_{n-1\hspace{2pt}l+2}$, and hence, for instance,
inferences of stellar ages. Here $\nu_{nl}$ is the frequency of a
mode with spherical degree $l$ and radial order $n$. The use of
$\delta\nu_{l \hspace{1pt} l+2}(n)$ for the purpose of solar-age
determination has been presented in earlier works (e.g. Dziembowski
et al.~1999; Bonanno et al.~2002). It has also been shown by several
authors (e.g. Roxburgh \& Vorontsov 2003; Ot\'{i} Floranes et
al.~2005; Christensen-Dalsgaard 2009) that the frequency separation
ratios, $r_{l \hspace{1pt} l+2
}(n)=(\nu_{nl}-\nu_{n-1\hspace{1pt}l+2})/(\nu_{nl}-\nu_{n-1\hspace{1pt}l})$,
are not as sensitive to the near-surface layers.
Christensen-Dalsgaard (2009) provides a comparison between the use
of $\delta\nu_{l \hspace{1pt} l+2}(n)$ and $r_{l \hspace{1pt} l+2
}(n)$. Here, we present our preliminary results on comparing
different ways of using the seismic data for determination of the
solar age, including the application of near-surface correction.

\fussy

\vspace{-8pt}
\section{Tools and methods}

\sloppy

We computed a series of standard solar models, using the stellar
evolution codes, ASTEC (Christensen-Dalsgaard 2008a) and GARSTEC
(Weiss \& Schlattl 2008), with different ages while keeping the
luminosity, and the surface value of $Z/X$ fixed to 3.846 $\times$
10$^{33}$ erg s$^{-1}$, and 0.0245 (Grevesse \& Noels 1993),
respectively. Here $Z$ is the mass fraction of the elements heavier
than helium, and $X$ is that of the hydrogen. Models were computed
with the OPAL equation of state (Rogers \& Nayfonov 2002), OPAL
opacities (Iglesias \& Rogers 1996) together with the
low-temperature opacities from Alexander \& Ferguson 1994, and using
the Adelberger et al.~(1998) or NACRE (Angulo et al.~1999) nuclear
reaction rates. We considered both the commonly used value $R_1 =
6.9599\times10^{10}$ cm (Auwers 1891) of the solar radius, and the
value $R_2 = 6.9551\times10^{10}$ cm found by Brown \&
Christensen-Dalsgaard (1998). Models were computed both starting at
the zero age main sequence (ZAMS) and including pre-main-sequence
(PMS) evolution. We calculated the frequencies of our models using
ADIPLS (Christensen-Dalsgaard 2008b), and compared the small
frequency separation, $\delta\nu_{02}(n)$, and $r_{02}(n)$ of the
models with those of the Sun (BISON, Chaplin et al.~2007). We
selected the best model minimizing the following $\chi^{2}$, adapted
for the separation ratios, $r_{02}$(n), when relevant:

\fussy

\begin{table*}[htbf]
\begin{center}
 \caption{Ages of the best models}
\begin{tabular}{@{}|c|c|c|c|c|c|}
\cline{3-6}
 \multicolumn{2}{c|}{}  & \multicolumn{2}{c|}{\multirow{2}{*}{$R_{1}$ = 6.9599$\times$10$^{10}$cm}}& \multicolumn{2}{c|}{\multirow{2}{*}{$R_{2}$ = 6.9551$\times$10$^{10}$cm}} \\
 \multicolumn{2}{c|}{}  &  \multicolumn{2}{c|}{}& \multicolumn{2}{|c|}{} \\
\hline
Evolution code +&Seismic property&Age$^{\footnotesize{*}}$ (before & Age$^{\footnotesize{*}}$ (after&Age$^{\footnotesize{*}}$ (before&Age$^{\footnotesize{*}}$ (after\\
Nuc. reaction rates:&used in $\chi^2$&correction)(Gyr)&correction)(Gyr)&correction)(Gyr)&correction)(Gyr)\\
\hline
ASTEC + &$\delta\nu_{02}(n)$&4.58$\pm$0.05&4.56$\pm$0.05&4.60$\pm$0.06&4.57$\pm$0.06\\
\cline{2-6}
Angulo et al.~(1999)&$r_{02}(n)$&4.55$\pm$0.04&4.55$\pm$0.04&4.55$\pm$0.05&4.55$\pm$0.05\\
\hline
ASTEC +&$\delta\nu_{02}(n)$&4.57$\pm$0.05&4.55$\pm$0.05&4.59$\pm$0.05&4.56$\pm$0.06\\
\cline{2-6}
Adelberger et al.~(1998)&$r_{02}(n)$&4.54$\pm$0.05&4.54$\pm$0.05&4.54$\pm$0.05&4.54$\pm$0.05\\
\hline
&\multirow{2}{*}{$\delta\nu_{02}(n)$}&\multirow{2}{*}{4.60$\pm$0.08}&\multirow{2}{*}{4.56$\pm$0.08}&4.61$\pm$0.08&4.56$\pm$0.08\\
GARSTEC +&&&&\scriptsize{(PMS: 4.67$\pm$0.08)}&\scriptsize{(PMS: 4.62$\pm$0.08)}\\
\cline{2-6}
Angulo et al.~(1999)&\multirow{2}{*}{$r_{02}(n)$}&\multirow{2}{*}{4.57$\pm$0.08}&\multirow{2}{*}{4.57$\pm$0.08}&4.57$\pm$0.08&4.57$\pm$0.08\\
&&&&\scriptsize{(PMS: 4.62$\pm$0.08)}&\scriptsize{(PMS: 4.62$\pm$0.08)}\\
\hline
\end{tabular}
\end{center}
\footnotesize{ $^{*}$ Evolution starts with ZAMS (Zero Age Main
Sequence) unless otherwise stated.}
\end{table*}

$${\footnotesize
\chi^{2}(\delta\nu_{02})=\frac{1}{N-1}\sum_{n}\frac{[\delta\nu_{02}^{\rm{obs}}(n)-\delta\nu_{02}^{\rm{model}}(n)]^2}{\sigma[\delta\nu_{02}(n)]^2}},
$$where $N$ is the number of modes, $\delta\nu_{02}^{\rm{obs}}(n)$, and
$\delta\nu_{02}^{\rm{model}}(n)$ are the small frequency separations
from the observations and the models; $\sigma[\delta\nu_{02}(n)]$
represents the uncertainty in the observed $\delta\nu_{02}(n)$. We
repeated this comparison for $\delta\nu_{02}(n)$ and $r_{02}(n)$ of
the surface-corrected frequencies (cf.\ Eq.\ 1).

\begin{figure}[h]
\centering 
\includegraphics[scale=0.30]{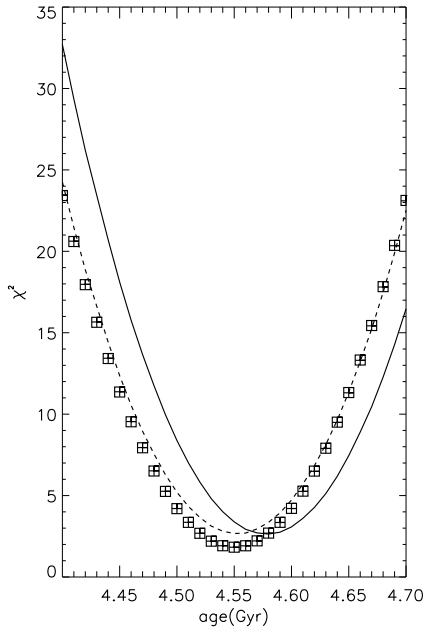}
\caption{The results with $R_{1}$, using NACRE (Angulo et al.~1999).
We present the $\chi^2$ obtained using $\delta\nu_{02}(n)$ of the
uncorrected model frequencies (solid line), $\delta\nu_{02}(n)$ of
the corrected model frequencies (dashed line), $r_{02}(n)$ of the
uncorrected model frequencies (squares), and $r_{02}(n)$ of the
corrected model frequencies (crosses).} \label{label_for_figure}
\end{figure}

\section{Results and discussion}

The results of the age determinations using different input, as
described in Section 2, are summarized in Table 1, and one of the
cases is illustrated in Fig. 1.

The ages of our best models found using the two different evolution
codes are consistent within the error limits, differing by 0.7\% at
most, and they are also compatible with the values in the literature
obtained by employing small frequency separations: 4.57 $\pm$ 0.11
Gyr (Bonanno et al.~2002), and 4.66 $\pm$ 0.11 Gyr (Dziembowski et
al.~1999). Moreover, our two sets of results agree with the
meteoritic age, 4.563 Gyr $<$ t $<$ 4.576 Gyr (Wasserburg, in
Bahcall \& Pinsonneault 1995), within the uncertainty limits.

In principle we can determine the duration of the PMS evolution, up
to the ZAMS, by calibrating the models using PMS evolution to get
the meteoritic age of the Sun (see, e.g. Morel et al.~2000, for a
discussion on the definition of ZAMS). Our preliminary analysis
suggests that the PMS evolution for the Sun is 0.05$\pm$0.11 Gyr in
order to be consistent with the meteoritic age. A more precise and
detailed analysis is obviously needed for defining ZAMS.

The results obtained using separation ratios are consistent with
those obtained using the small frequency separation of the corrected
frequencies. However, since the application of the surface
correction has no significant effect on the age determined by the
separation ratios, using these ratios yields a more robust age
determination.

\acknowledgements This work was supported by the European Helio- and
Asteroseismology Network (HELAS), a major international
collaboration funded by the European Commission's Sixth Framework
Programme. GD and JC-D acknowledge financial support from the Danish
Natural Science Research Council.


\end{document}